\documentclass[epj]{webofc}
\usepackage[utf8]{inputenc}
\usepackage[varg]{txfonts}   % Web of Conferences font
\usepackage{booktabs}
\usepackage{xcolor}
\usepackage{float}
\usepackage{dsfont}
\definecolor{darkred}{rgb}{0.4,0.0,0.0}
\definecolor{darkgreen}{rgb}{0.0,0.4,0.0}
\definecolor{darkblue}{rgb}{0.0,0.0,0.4}
\usepackage[bookmarks,linktocpage,colorlinks,
    linkcolor = darkred,
    urlcolor  = darkblue,
    citecolor = darkgreen]{hyperref}
\usepackage{subcaption}
\wocname{EPJ Web of Conferences}
\woctitle{Lattice2017}
\begin{document}
%%%%%%%%%%%%%%%%%%%%%%%%%%%%%%%%%%%%%%%%%%%%%%%%%%%%%%%%%%%%%%%%%
%
\selectlanguage{english}
%----------------------------------------------------------------------------
\title{%
New techniques and results for worldline simulations of lattice field theories
}
%----------------------------------------------------------------------------
\author{%
\firstname{Mario}  \lastname{Giuliani}\thanks{Speaker, \email{mario.giuliani@uni-graz.at}},
\firstname{Oliver}  \lastname{Orasch}\thanks{Speaker, \email{oliver.orasch@edu.uni-graz.at}}
\firstname{Christof}  \lastname{Gattringer}\thanks{\email{christof.gattringer@uni-graz.at}},
% etc.
}
%----------------------------------------------------------------------------
\institute{%
Institut f\"ur Physik, Universit\"at Graz, 8010 Graz, Austria 
}
%----------------------------------------------------------------------------
\abstract{%
We use the complex $\phi^4$ field at finite density as a model system for developing further techniques based on 
worldline formulations of lattice field theories. More specifically we: 
1) Discuss new variants of the worm algorithm for updating the $\phi^4$ theory and related systems with site weights. 
2) Explore the possibility of canonical simulations in the worldline formulation. 
3) Study the connection of 2-particle condensation at low temperature to scattering parameters of the theory.
}
%----------------------------------------------------------------------------
\maketitle
%----------------------------------------------------------------------------
\section{Introduction}\label{intro}

In recent years we have seen a fast development of dualization techniques for lattice field theories which is largely motivated by 
exploring new approaches for overcoming complex action problems for simulations at finite chemical potential (see, e.g., the reviews 
\cite{Chandrasekharan:2008gp,deForcrand:2010ys,Wolff:2010zu,Gattringer:2014nxa,Gattringer:2016kco}). In a dual formulation
the lattice field theory is exactly rewritten in terms of new variables which correspond to worldlines for matter fields and worldsheets for
gauge fields. If the weights for the dual configurations are real and positive a Monte Carlo simulation can be done directly in terms
of the dual variables and the complex action problem is solved.

A simple but important lattice quantum field theory where worldline techniques are widely applicable is the complex $\phi^4$ field,
often also referred to as the ''relativistic Bose gas''. The first successful dual representation was discussed by Endres in 
\cite{endres1,endres2} and since then the complex $\phi^4$ field has been used as a model system for many important 
developments of worldline techniques 
\cite{berlin1,berlin2,berlin3,berlin4,berlin5,berlin6,phi4_1,phi4_2,phi4_3,worm_phi4,canonical_phi4}, as
well as as a reference system for other approaches to overcome the complex action problem (see, e.g., 
\cite{Aarts:2008wh,Aarts:2009hn,Akerlund:2014mea,Alexandru:2016san,Cristoforetti:2013wha,Bongiovanni:2016jdj} in this context).

In this contribution we continue the development and testing of new worldline techniques in the complex $\phi^4$ model 
system. More specifically we revisit the problem of optimizing the worm algorithm, explore the possibility of canonical worldline 
simulations and study the relation of the 2-particle condensation thresholds at low temperatures to scattering data of the theory. 
All three topics, improved worm algorithms, canonical worldline simulations and the relation of scattering data to low temperature 
condensation are methods and phenomena with implications beyond the simple $\phi^4$ theory which here is merely 
considered for developing the techniques and physical ideas.

\section{Worldline representation of the complex $\phi^4$ field}

As already outlined in the introduction, we here discuss techniques and results developed for the worldline representation of 
the complex $\phi^4$ field. Although the worldline representation is well known, we briefly summarize it in this section to 
make clear its form and our notations. 

The lattice action for the conventional representation of the complex $\phi^4$ field in $d$-dimensions is,
\begin{equation}
S[\phi]  \; = \; 
\sum_{x \in \Lambda} \left( \eta \, |\phi _{x}|^2 \; + \; \lambda \, |\phi _{x}|^4 \; - \; \sum_{\nu = 1}^{d} 
\left[ e^{\,\mu \delta_{\nu,d}}\phi _{x}^{\ast} \phi _{x+\hat{\nu}} + e^{\,-\mu \delta_{\nu,d}}\phi _{x}^{\ast} \phi _{x-\hat{\nu}} \right]\right) \; .
\label{action_conventional}
\end{equation}
In the conventional representation the degrees of freedom are the complex valued fields $\phi_x \in \mathds{C}$ 
assigned to the sites $x$ of a $d$-dimensional lattice $\Lambda$ with periodic boundary 
conditions. The spatial extent is denoted by $N_s$ and the temporal extent by $N_t$. The latter corresponds to the inverse temperature 
$\beta$ in lattice units, i.e., $\beta = N_t$. Here we consider the cases of $d = 2$ and $d=4$, i.e., we work on lattices with volumes 
$V = N_s \times N_t$ and $V = N_s^3 \times N_t$. The bare mass $m$ enters via the parameter $\eta \equiv 2d + m^{2}$ 
and $\lambda$ denotes the quartic self-interaction.
The chemical potential $\mu$ gives a different weight for forward and backward propagation in the Euclidean time direction 
($\nu = d$). The grand canonical partition sum is given by $Z   = \int D[\phi] e^{-S[\phi]}$, where in the path integral we integrate 
with a measure that is the product of complex integrals, $\int \! D[\phi] = \prod_x \int_\mathds{C} d \phi_x/2\pi$.
 
\begin{figure}
\begin{center}
\includegraphics[height=65mm,clip]{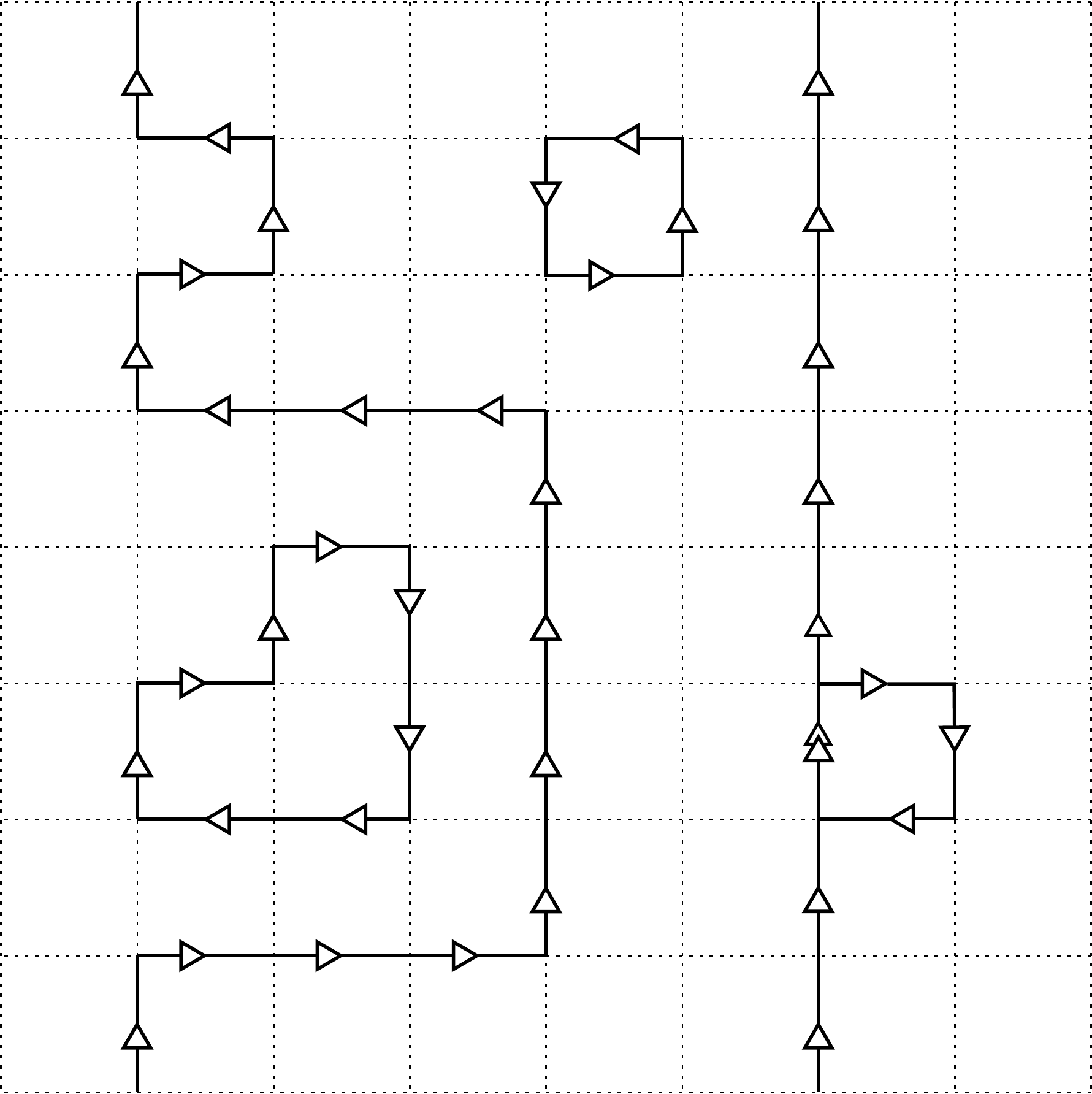}
\end{center}
\caption{Example of a configuration of dual variables. The dual variables correspond to conserved flux and thus must form 
worldlines. Their temporal winding number corresponds to the net particle number of the configuration, i.e., the example configuration we show has a net-particle number of $+2$.} 
\label{fluxlines}
\end{figure}

It is obvious, that for non-zero $\mu$ the action has a non-vanishing imaginary part and thus we face a complex action problem. However,
this complex action problem may be overcome by exactly transforming the partition sum to a dual representation (see, e.g., \cite{phi4_1} 
for the derivation of the form we use here). In the dual representation the grand canonical partition function is a sum $\sum_{\{k\}}$ over 
configurations of flux variables  $k_{x,\nu} \in \mathds{Z}$ assigned to the links of the lattice,
\begin{equation}
Z   \; = \; \sum_{\{k\}}  e^{\, \mu \, \beta \, \omega[k]} \; B[k] \; \prod_{x} \delta \left(\vec{\nabla} \cdot \vec{k}_{x} \right) \; .
\label{worldlineZ}
\end{equation}
The flux variables $k_{x,\nu}$ are subject to constraints which in the dual form (\ref{worldlineZ}) are represented by the product 
$\prod_{x \in \Lambda} \delta (\vec{\nabla} \cdot \vec{k}_{x} )$ 
over all sites $x$ of the lattice, where by $\delta(j) \equiv \delta_{j,0}$
we denote the Kronecker Delta. $\vec{\nabla} \cdot \vec{k}_{x}$ is the lattice form of the divergence of $k_{x,\nu}$ given by
$\vec{\nabla} \cdot \vec{k}_{x}  \equiv   \sum_{\nu}(k_{x,\nu} - k_{x-\hat{\nu},\nu})$. The zero-divergence constraint implies
that the net flux of $k_{x,\nu}$ vanishes at every site $x$ and consequently the admissible configurations of the $k_{x,\nu}$ 
have the form of worldlines of conserved flux. In Fig.~\ref{fluxlines} we show an example of an admissible configuration 
of $k$-flux.

From the figure it is obvious that the worldlines can wind around the spatial and temporal directions due
to the periodic boundary conditions. In the example of Fig.~\ref{fluxlines} we show an example of a worldline configuration
with a temporal net winding number of $+2$ (the vertical direction is time in the illustration).  We denote the temporal winding 
number of a configuration of $k$-worldlines by $\omega[k]$, and this temporal winding number is exactly the geometrical 
quantity the chemical potential $\mu$ couples to.  $\omega[k]$ appears in the form of $e^{\, \mu \, \beta \, \omega[k]}$, such that 
from the canonical form of the coupling of particle number $N$ and  chemical potential $\mu$
we can identify the temporal winding number $\omega[k]$ with the net particle number $N$. Note that this form of the net particle
number as a topological quantity (the temporal winding number) allows one to uniquely assign an exact integer 
net particle number to any given configuration 
of the worldline variables. Thus the worldline formulation is very suitable for canonical simulations which we will address in Section 4.
This aspect is different from the conventional representation where one has to work with a discretization of
the continuum Noether charge, which does not give an integer result.   
 
In addition to the chemical potential factor $e^{\, \mu \, \beta \, \omega[k]}$ and the Kronecker constraints, each configuration of
$k$-fluxes also comes with a weight factor $B[k]$. This factor is itself a sum over configurations $\sum_{\{a\}}$ of integer valued 
auxiliary link variables $a_{x,\nu} \in \mathds{N}_0$ and is given by 
\begin{eqnarray}
&& \hspace{30mm}
B[k] \; = \; \sum_{\{a\}} \, \prod_{x,\nu}\frac{1}{(a_{x,\nu}+|k_{x,\nu}|)! \, a_{x,\nu}!} \,  \prod_{x} I(s_x) \qquad \mbox{with} 
\label{weights} \\
&& 
I(s_x) \; = \; \int_{0}^{\infty} \!\! d r \; r^{\, s_x + 1} \, e^{\, -\eta \, r^2 \, - \, \lambda \, r^4} \quad \mbox{and} \quad
s_x \; = \; \sum_{\nu}\Big[|k_{x,\nu}| +  |k_{x-\hat{\nu}}| + 2(a_{x,\nu} +  a_{x-\hat{\nu}})\Big] \; .  \nonumber
\end{eqnarray}
Obviously the integrals $I(s_x)$ come from integrating out the radial degrees of freedom of the original field variables at site $x$.  
The argument $s_x$ is a non-negative integer combination of the auxiliary variables and the moduli of all $k$-fluxes that run
through $x$. For a numerical simulation the integrals $I(s_x)$ can be pre-calculated and stored for  
sufficiently many values of the arguments $s_x \in \mathds{N}_0$.

Since all weights, those in the dual partition sum (\ref{worldlineZ}), as well as those in the weight factor $B[k]$ are real and 
positive we can analyze the system in a Monte Carlo simulation using the flux variables $k_{x,\nu}$ and the auxiliary variables $a_{x,\nu}$. 
Thus the dual representation solves the complex action problem. 

It is straightforward to also transform observables to their dual representation. The observables we are interested in here are 
derivatives of the free energy density $f \, \equiv \, -  \ln Z \, / \, (N_s^{d-1} \beta) \, = \, -  \ln Z / V$ 
with respect to the parameters. These derivatives can then be evaluated also for the dual formulation
and the result is the dual representation of the observables. For the examples of the particle number density $n = N\, / \, (N_s^{d-1})$ 
and the field expectation value $|\phi |^2$ the corresponding dual representations are 
\begin{equation}
\langle n \rangle  = - \frac{\partial f}{\partial \mu}  =    \frac{\langle \omega \rangle }{N_s^{d-1}} \; , \; \; 
\langle |\phi |^2 \rangle =  \frac{\partial f}{\partial \eta} =  \frac{1}{V} \!
\left\langle \sum_{x} \!\frac{I(s_x\!+\!2)}{I(s_x)} \right\rangle \; .
\label{observables}
\end{equation}
The vacuum expectation values on the right hand sides are now understood in the dual representation.

\section{New concepts for Monte Carlo simulations with worm algorithms}

As we have seen in the previous section the variables of the dual representation are subject to constraints. In particular the 
flux variables $k_{x,\nu}$ must obey the zero-divergence condition at every site $x$ and consequently the admissible 
configurations are worldlines of conserved flux. For this type of systems the Prokofev-Svistunov worm algorithm \cite{worm}
provides a powerful strategy which takes into account the constraints in a natural way. 

However, the $\phi^4$ field has a feature that goes beyond simpler cases such as the Ising or O(2) models where the 
original degrees of freedom are group valued. The scalar $\phi_x$ field not only has a complex phase in U(1), 
but also a radial degree of freedom, i.e., the modulus $|\phi_x|$. Obviously the modulus is essential for the physics of the
theory since it appears in the mass term and the quartic term whose interplay determines whether the model is in the massive
or in the Higgs phase. As can be seen from the form of the weights (\ref{weights}) of the dual representation, the radial degrees 
of freedom give rise to the specific weight factors $I(s_x)$ that live on the sites $x$ and via  
$s_x  = \sum_{\nu}[|k_{x,\nu}| +  |k_{x-\hat{\nu}}| + 2(a_{x,\nu} +  a_{x-\hat{\nu}})]$ depend on the fluxes on all links that 
attach to $x$. This makes the worm algorithm more involved, since at every link $(x,\nu)$ which 
the worm adds to its contour we do not know the final weight at the endpoint $x + \hat{\nu}$ of the link, 
since this weight is fully determined only after the subsequent step of the worm. This feature leads to an imbalance in particular 
for the initial and final steps of the worm that may lead to an inefficient algorithm in some parameter range, unless
suitable additional strategies are invoked which we discuss here (see also \cite{worm_phi4}).

Before we come to addressing the new ideas for worm algorithms for systems with additional site weights, let us point out that 
the site weights might also come from different sources, in particular when the Haar measure from a non-abelian group 
enters the weight factors (see \cite{principal_chiral_1,principal_chiral_2,acc_su2} for examples). To have a more general 
notation that covers all such cases we consider a system of a conserved flux $k_{x,\nu}$ with partition sum
(the auxiliary variables $a_{x,\nu}$ of (\ref{weights}) are here considered as a background field that can be updated 
with a conventional local Monte Carlo)
\begin{equation}
Z  \; = \; \sum_{\{k\}} \! \left[ \prod_{x} \delta \left(\vec{\nabla} \cdot \vec{k}_{x} \right) \right]  
\left[ \, \prod_{x,\nu} L_{x,\nu} (k_{x,\nu}) \right] 
\left[ \prod_x S_x( \{k_{x,\bullet} \}) \right] 
\; .
\label{Zflux}
\end{equation}
Here we allow for link weights $L_{x,\nu} (k_{x,\nu})$ that depend on the value of $k_{x,\nu}$ on that particular link as well as
the position of the link such that also the spatial dependence from the background field
$a_{x,\nu}$ of (\ref{weights}) is taken into account. We also include site weights $S_x( \{k_{x,\bullet} \})$ that depend on the
fluxes $k_{x,\bullet}$ on all links that attach to $x$. It is easy to see that the dual formulation (\ref{worldlineZ}), (\ref{weights}) 
of the charged $\phi^4$ field is of the form (\ref{Zflux}), but also other theories such as the worldline form of the principal chiral
model \cite{principal_chiral_1,principal_chiral_2}.

For systems of the form (\ref{Zflux}) we now present two new strategies for efficient worm algorithms. 
The first one is the introduction of an amplitude parameter $A$ that can be used to tune the length of the worms and thus 
the average number of fluxes changed by every completed worm. The second idea is to impose an even-odd decomposition of the lattice  
and to use different steps for moves of the worm that lead from an even to an odd site and moves that lead from an odd to an even site. 

\vskip3mm
\noindent
In general a worm consists of three parts: (we here use the convention $k_{x,-\nu}  \equiv k_{x-\hat\nu,\nu}$)

\begin{itemize}

\item[-] Worm start: The worm randomly chooses a flux increment $\Delta \in \{-1,1\}$, a starting site $x_0$ and
an initial direction $\nu_0 \in \{\pm 1\, ... \, \pm d\}$. A defect that violates the flux conservation constraints is inserted by proposing to 
change $k_{x_0,\nu_0} \rightarrow k_{x_0,\nu_0} + \text{sign}(\nu_0)\Delta$ which is accepted with a Metropolis 
decision\footnote{The Metropolis propability $P$ for a accepting a proposed move is related to the Metropolis ratio $\rho$
via $P =$ min $\{\rho,1\}$.} with Metropolis ratio $\rho_{x_0,\nu_0}^{\,S}$. Upon acceptance the worm head proceeds to 
$x_1 = x_0 + \hat{\nu}_0$.

\item[-]Worm propagation: Departing from the current site $x_j$ of the head of the worm a 
new direction $\nu_j \in \{\pm 1\, ... \, \pm d\}$ is chosen randomly and the change 
$k_{x_j,\nu_j} \rightarrow k_{x_j,\nu_j} + \text{sign}(\nu_j)\Delta$ is proposed. The propagation steps are accepted with a
different Metropolis ratio $\rho_{x_j,\nu_j}^{\,P}$. The algorithm keeps proposing new directions until one is accepted and the 
head of the worm proceeds to $x_{j+1} = x_j + \hat{\nu}_j$.

\item[-]Worm closing: In case the site $x_{j+1} = x_j + \hat{\nu}_j$ that would be reached in an accepted propagation step coincides with 
the starting site of the worm, i.e., $x_{j+1} = x_0$, the worm uses a different Metropolis ratio $\rho_{x_{j},\nu_{j}}^{\,C}$ for accepting 
that step. If accepted, the worm terminates, thus healing the initially induced defect, 
and the new configuration again obeys all flux conservation constraints.
\end{itemize}

\noindent
We will show that the following assignments for the Metropolis ratios 
$\rho_{x_0,\nu_0}^{\,S}$ $\rho_{x_j,\nu_j}^{\,P}$ and $\rho_{x_{n-1},\nu_{n-1}}^{\,C}$ give rise to a worm algorithm that obeys 
detailed balance for systems of the form (\ref{Zflux}):

\begin{eqnarray} 
&& \hspace*{-10mm}\rho_{x_0,\nu_0}^{\,S} \; = \; \frac{A}{
S_{x_0}( \{k_{x_0,\bullet} \}) \; S_{x_1}( \{k_{x_1,\bullet} \})} \;
\frac{ L_{x_0,\nu_0}(k_{x_0,\nu_0}^{trial})}{L_{x_0,\nu_0}(k_{x_0,\nu_0})} \quad ,
\quad
\rho_{x_j,\nu_j}^{\,P}  =   \frac{
S_{x_j} ( \{ k^{new}_{x_j,\bullet} \}) }{ 
S_{x_{j+1}}( \{k_{x_{j+1},\bullet} \}) } \;
\frac{ L_{x_j,\nu_j}(k_{x_j,\nu_j}^{trial})}{L_{x_j,\nu_j}(k_{x_j,\nu_j})} \; ,
\nonumber \\
&& \hspace{15mm} \rho_{x_{n-1},\nu_{n-1}}^{\,C}  =  \frac{
S_{x_{n-1}} ( \{ k^{new}_{x_{n-1},\bullet}  \}) \; 
S_{x_0}       ( \{ k^{new}_{x_0,\bullet}  \}) } {A} \;
\frac{ L_{x_{n-1},\nu_{n-1}}(k_{x_{n-1},\nu_{n-1}}^{trial})}{L_{x_{n-1},\nu_{n-1}}(k_{x_{n-1},\nu_{n-1}})} \; ,
\label{rhos}
\end{eqnarray}
where for the Metropolis ratio of the closing step we have already assumed that the worm has length $n$, i.e., the closing step leads
from $x_{n-1}$ to $x_n = x_0$. The role of the amplitude $A \in \mathds{R}_+$ will be discussed later.
For the subsequent proof of detailed balance we introduce the notation 
$w \; = \; (  \Delta,  x_0, \nu_0, \nu_1, \, ... \, \nu_{n-1})$ for a worm, which denotes all steps of the worm in a compact form.  

In order to establish that the worm generates the correct probability distribution $W(C_k)$ for flux configurations $C_k$, we show the 
sufficient (but not necessary) detailed balance condition
\begin{equation}
\frac{P(C_k \rightarrow \widetilde{C}_k)}{P(\widetilde{C}_k \rightarrow C_k)} \; = \; \frac{W(\widetilde{C}_k)}{W(C_k)} \; ,
\label{detailedbalance}
\end{equation}
where $P(C_k \rightarrow \widetilde{C}_k)$ is the probability to go from configuration $C_k$ to $\widetilde{C}_k$. Note that here
$\widetilde{C}_k$ denotes the new configuration after the worm has closed, since intermediate steps of the worm do not correspond 
to admissible configurations that obey the constraints. The configuration $\widetilde{C}_k$ differs from $C_k$ by a set of links where the
flux variables $k_{x,\nu}$ were changed. It is important to note that there is an infinity of loops that give rise to that particular
change of flux. These worms can differ by their starting point $x_0$, their flux increment $\Delta$ but also by the sequence of directions
$\nu_j$, since the worm can produce dangling ends where it retraces previous steps and thus reverts changes of flux. 
Consequently we need to write the detailed balance condition (\ref{detailedbalance}) in the form
\begin{equation}
\frac{\sum_{\{w\}} \,\, P_w \, (C_k \rightarrow \widetilde{C}_k)}{
\sum_{\{w^\star\}} P_{w^\star} (\widetilde{C}_k \rightarrow C_k)} \; = \; \frac{W(\widetilde{C}_k)}{W(C_k)} \; ,
\label{detailedbalance2}
\end{equation}
where the transition probabilities $P(C_k \rightarrow \widetilde{C}_k)$ are written as sums over the set $\{w\}$ of all 
worms that lead from $C_k$ to $\widetilde{C}_k$, and by $\{w^\star\}$ we denote the corresponding set of worms 
that go from $\widetilde{C}_k$ to $C_k$.  The transition probability for an individual worm $w$ is denoted by 
$P_w \, (C_k \rightarrow \widetilde{C}_k)$. A sufficient solution of (\ref{detailedbalance2}) is to show the existence of a
bijective mapping $f$ between $\{w\}$ and $\{w^\star\}$ such that each pair of a worm $w$ and the inverse worm $w^{-1} = f(w)$  
individually obey detailed balance:
\begin{equation}
\exists \; \mbox{bijective} \; f \, : \,  \{w\} \overset{f}{\leftrightarrow} \{w^\star\} \; \forall \; w \; : \, 
\frac{P_w(C_k \rightarrow \widetilde{C}_k)}{P_{f(w)}(\widetilde{C}_k \rightarrow C_k)} =  \frac{W(\widetilde{C}_k)}{W(C_k)}  \; .
\label{detailedbalance3}
\end{equation}
We can identify a suitable bijective mapping $f$ by defining the inverse worm $w^{-1} = f(w)$ as the worm with the same flux 
increment $\Delta$ and the same starting point $x_0$, but reversed orientation:
\begin{equation}
w \; = \; ( \Delta,  x_0, \nu_0 , \nu_1, \, ... \, \nu_{n-1}) \quad \rightarrow \quad 
w^{-1} \; \equiv \; ( \Delta, x_0,-\nu_{n-1}, -\nu_{n-2}, \, ... \, - \nu_1, -\nu_0) \; .
\label{inverseworm}
\end{equation}
Obviously $w^{-1}$ transforms $\widetilde{C}_k$ back to $C_k$. It is important to stress that with the choice (\ref{inverseworm}) 
at every site of the worm path, 
$w$ and $w^{-1}$ encounter the same values of the variables $k_{x,\nu}$ such that the probabilities for accepting the steps
are computed in the same background of flux variables. 

The transition probabilities $P_w(\widetilde{C}_k \rightarrow C_k)$ in (\ref{detailedbalance3}) are now 
obtained as the product of the probabilities for the individual steps in the worm $w$. For the three different parts, 
starting, propagating and closing the corresponding probabilities are 
$P^S_{x_0,\nu_0} = \mbox{min} \, \{ \rho^{S}_{x_0,\nu_0} , 1 \} / 2dV$, 
$P^P_{x_j,\nu_j} =  \mbox{min} \, \{  \rho^{P}_{x_j,\nu_j}, 1 \} / N_{x_j} $ and 
$P^C_{x_{n-1},\nu_{n-1}} = \mbox{min} \, \{  \rho^{C}_{x_{n-1},\nu_{n-1}}, 1  \} / N_{x_{n-1}} $, 
where we have normalized the probabilities  by summing over all possible choices at each step. For the starting step this is
the trivial factor $2dV$ for choosing the first link and its orientation, while for the propagating and closing steps the normalization reads
\begin{equation}
N_{x_j} \, = \; \sum_{\sigma = \pm 1}^{\pm d} \Big[ \mbox{min} \, \Big\{  \rho^{P}_{x_j,\sigma}, 1 \Big\} 
[ 1 -  \delta_{x_j+\hat{\sigma},x_0} ]  \; + \; \mbox{min} \, \Big\{  \rho^{C}_{x_j,\sigma}, 1 \Big\} \delta_{x_j+\hat{\sigma},x_0} \Big]  \; .
\end{equation}
Thus the transition probability of a worm path reads
$P_w (C_k \rightarrow \widetilde{C}_k)  =  P^S_{x_0,\nu_0} \left( \prod_{j=1}^{n-2} P^P_{x_j,\nu_j} \right) \, 
P^C_{x_{n-1},\nu_{n-1}}$ and the sufficient condition (\ref{detailedbalance3}) for detailed balance turns into
\begin{equation}
\frac{P_w (C_k \rightarrow \widetilde{C}_k)}{P_{w^{-1}} (\widetilde{C}_k \rightarrow C_k)} \; = \; 
\frac{ P^S_{x_0,\nu_0} \; \left( \prod_{j=1}^{n-2} P^P_{x_j,\nu_j} \right) \; \, P^C_{x_{n-1},\nu_{n-1}} }
{\tilde P^S_{x_0,-\nu_{n-1}} \; \left( \prod_{j={n-1}}^{2} \tilde P^P_{x_j,-\nu_{j-1}} \right) \; \, \tilde P^{\,C}_{x_{1},-\nu_{0}} }  
\; = \; \frac{W(\widetilde{C}_k)}{W(C_k)} \; ,
\label{ratioworm2}
\end{equation}
where the probabilities of the individual steps of the inverse worm are marked with a twiddle.
We stress again that with our choice of $w^{-1}$ at each step the worm $w$ and its inverse $w^{-1}$ make their decisions 
for the next direction in the same background of flux variables such that all normalizations cancel 
between the numerator and the denominator. Thus (\ref{ratioworm2}) reduces further to
\begin{equation}
\frac{P_w (C_k \rightarrow \widetilde{C}_k)}{P_{w^{-1}} (\widetilde{C}_k \rightarrow C_k)} \; = \; 
\frac{ \, \mbox{min} \{\rho^S_{x_0,\nu_0} , 1 \} }{ \, \mbox{min} \{ \tilde \rho^{\,C}_{x_{1},-\nu_{0}} , 1 \} } \!
\left( \prod_{j=1}^{n-2} 
\frac{\, \mbox{min} \{ \rho^P_{x_j,\nu_j}, 1 \} } { \, \mbox{min} \{ \tilde \rho^P_{x_{j+1},-\nu_{j}},1\} } \! \right) \! 
\frac{  \, \mbox{min} \{\rho^C_{x_{n-1},\nu_{n-1}} , 1 \} } { \, \mbox{min} \{ \tilde\rho^S_{x_0,-\nu_{n-1}}, 1 \} }
\; ,
\label{ratioworm3}
\end{equation}
where we again use twiddles to denote the Metropolis ratios of the inverse worm. 
We also have reordered the factors such that we match the steps of $w$ and $w^{-1}$ at the same link in the same factor. 
Inspecting our chosen Metropolis ratios (\ref{rhos}) one easily shows the  properties,
\begin{equation}
\tilde \rho^{\,C}_{x_{1},-\nu_{0}} = \left( \rho^S_{x_0,\nu_0}  \right)^{-1} \; , \; \; 
\tilde \rho^P_{x_{j+1},-\nu_{j}} = \left( \rho^P_{x_j,\nu_j} \right)^{-1} \; , \; \;  
\tilde \rho^S_{x_0,-\nu_{n-1}} = \left( \rho^C_{x_{n-1},\nu_{n-1}}\right)^{-1} .
\label{rhoprops}
\end{equation}
Finally we make use of the identity $\mbox{min} \{ \rho, 1\} / \mbox{min} \{ \rho^{-1}, 1\} = \rho$ for $\rho > 0$ and find
\begin{equation}
\frac{P_w (C_k \rightarrow \widetilde{C}_k)}{P_{w^{-1}} (\widetilde{C}_k \rightarrow C_k)} \; = \; 
\rho^S_{x_0,\nu_0} 
\left( \prod_{j=1}^{n-2} \rho^R_{x_j,\nu_j} \right) \rho^T_{x_{n-1},\nu_{n-1}} \; = \; \frac{W(\widetilde{C}_k)}{W(C_k)} \; .
\label{ratioworm4}
\end{equation}
This concludes our proof of detailed balance.

In our choice for the Metropolis ratios for the starting and closing steps in (\ref{rhos}) we have included the amplitude parameter 
$A \in \mathds{R}_+$, and obviously detailed balance holds for arbitrary choices of $A$. The usefulness of the parameter $A$ becomes
clear when inspecting the Metropolis ratios for the starting and closing steps. As already discussed, the fact that the site weights of 
(\ref{Zflux}) depend on all fluxes $k_{x,\nu}$ attached to a site $x$ implies that we cannot choose the Metropolis ratios for the starting 
and closing steps with the same number of site weights in the numerator and the denominator. As a consequence, in the starting ratio 
$\rho^S_{x_0,\nu_0}$ two factors of site weights $S_x( \{k_{x,\bullet} \})$ appear in the denominator. The typical numerical 
values of the site weights $S_x( \{k_{x,\bullet} \})$ will of course depend on the coupling values where the simulation is performed. 
For couplings where the site weights are large the form of $\rho^S_{x_0,\nu_0}$ will give rise to very small probabilities for accepting the 
starting step of a worm. In turn, the ratios $\rho^{C}_{x_{n-1},\nu_{n-1}}$ 
for the closing step, where the site weights appear only in the numerator (see (\ref{rhos})), will be larger
than 1, such that every closing attempt is accepted. Thus in such a region of coupling space we have worms that hardly start and 
then quickly terminate, i.e., are very short. The amplitude factor in (\ref{rhos}) can now be used to eliminate or at least 
milden the problem. Choosing $A$ proportional to the square of the characteristic site weight gives suitable starting and 
closing probabilities and thus worms of reasonable length, as was also confirmed in numerical tests documented in \cite{worm_phi4}.

Let us now come to the discussion of the second new worm strategy we are currently exploring. It is based on an even-odd
decomposition of the lattice and the fact that a worm alternates steps leading from an even to an odd site with steps that lead from an 
odd to an even site. The key idea is to use different Metropolis acceptance rates for even to odd (ETO) and odd to even (OTE) 
steps of the worm. For the inverse worm, at each link of the contour of the worm, 
the role of ETO and OTE are reversed such that in the detailed balance equation
for each link both ETO and OTE probabilities appear. This allows for an additional freedom in the choice of the acceptance probabilities
which can be explored to construct efficient worms for systems with the additional site weights we discussed.

As before the worm starts with randomly selecting a flux increment $\Delta  \in \{ \pm 1 \}$, a starting site $x_0$ and 
an initial direction $\nu_0 \in \{ \pm 1 \, ... \, \pm d \}$. Subsequent steps are determined by randomly choosing new directions 
$ \nu_j \in \{ \pm 1 \, ... \, \pm d \}$ and in every step the change of flux $k_{x_j,\nu_j} \rightarrow k_{x_j,\nu_j} + \text{sign}(\nu_j)\Delta$
is proposed. If the change is accepted, the head of the worm proceeds from $x_j$ to $x_j + \hat{\nu}_j$. Once the head reaches 
the starting site $x_0$ the worm terminates. Each step is accepted with a specific probability, and we use $P_{x_0,\nu_0}^{\,S}$ to denote
the probability for the starting step, $P_{x_j,\nu_j}^{\,P}$ for the intermediate propagation steps and 
$P_{x_{n-1},\nu_{n-1}}^{\,C}$ for accepting the closing step. 

As already announced, the acceptance probabilities are now chosen with respect to an even-odd pattern. Per definition the chosen 
staring site $x_0$ is labelled as even. Thus the first step of the worm is an ETO step followed by alternating OTE and ETO steps.
The final step is an OTE step, since the number of steps the worm needs for forming a closed contour is an even number (we 
stress that we assume that both $N_s$ and $N_t$ are even, such that this holds also for worms that wind around the boundaries). 
The probabilities for accepting the steps are now chosen as
\begin{equation} 
P_{x_0,\nu_0}^{\,S} \, = \, \frac{1}{2dV} \; , \;
{P_{x_j,\nu_j}^{\,P}} \, = \,  \frac{ \rho_{x_j,\nu_j}^{\,P} }{ N_{x_j} } \; \text{for OTE} \; , \;
{P_{x_j,\nu_j}^{\,P}} \, = \, \frac{ 1 - \delta_{x_j+\nu_j,x_{j-1}} }{ 2d-1 } \; \text{for ETO} \; , \; 
P_{x_{n-1},\nu_{n-1}}^{\,C} \, = \, \frac{ \rho_{x_{n-1},\nu_{n-1}}^{\,C} } {{ N_{x_{n-1}} }} \; ,
\label{newrhos}
\end{equation}
where the ratios $\rho_{x_j,\nu_j}^{\,P}$ and $\rho_{x_{n-1},\nu_{n-1}}^{\,C}$ are those of (\ref{rhos}). Note that the even-odd
decomposition allows one to eliminate backtracking for ETO steps, which is manifest in the choice of 
$P_{x_j,\nu_j}^{\,P}$ for ETO steps.

The proof of detailed balance proceeds in the same way as before, i.e., we again write the overall transition probability as
the sum over all worms that lead to the same change of flux and subsequently identify the inverse worm via the choice 
(\ref{inverseworm}).  The detailed balance condition (\ref{ratioworm2}) between a worm and its inverse now reads
\begin{equation}
\frac{P_w (C_k \rightarrow \widetilde{C}_k)}{P_{w^{-1}} (\widetilde{C}_k \rightarrow C_k)} \; = \; 
\frac{ \left( \prod_{j=0}^{\frac{n}{2}-2} P^P_{x_{2j+1},\nu_{2j+1}} \right)  \; P^C_{x_{n-1},\nu_{n-1}} }
{ \left( \prod_{j={\frac{n}{2}-1}}^{1} \tilde P^P_{x_{2j+1},-\nu_{2j}} \right)  \; \tilde P^{\,C}_{x_{1},-\nu_{0}} }  
\; = \; 
\frac{ \left( \prod_{j=0}^{\frac{n}{2}-2} \rho^P_{x_{2j+1},\nu_{2j+1}} \right) \; \rho^C_{x_{n-1},\nu_{n-1}} }
{ \left( \prod_{j={\frac{n}{2}-1}}^{1} \tilde \rho^P_{x_{2j+1},-\nu_{2j}} \right) \; \tilde \rho^{\,C}_{x_{1},-\nu_{0}} }  
\; = \; 
\frac{W(\widetilde{C}_k)}{W(C_k)} \; ,
\label{ratioworm2_new}
\end{equation} 
where in the first step we have already cancelled the constant (i.e., flux-independent)  
probabilities for all ETO steps, including the starting probability. In the 
second step we have inserted the explicit expressions for the probablities of the OTE steps from (\ref{newrhos}), and again the 
normalizations $N_{x_{j}}$ cancel for our choice of $w^{-1}$. We have already remarked that the ETO steps of $w^{-1}$ are the 
OTE steps of $w$. Thus in (\ref{ratioworm2_new}) indeed non-trivial flux dependent factors $\rho^P$ appear for all links
of the contour of the worm. We can complete the proof of detailed balance by using the relations (\ref{rhoprops}) to bring 
the factors from the denominator to the numerator and find,
\begin{equation}
\frac{P_w (C_k \rightarrow \widetilde{C}_k)}{P_{w^{-1}} (\widetilde{C}_k \rightarrow C_k)} \; = \; 
\rho^S_{x_0,\nu_0} 
\left( \prod_{j=1}^{n-2} \rho^R_{x_j,\nu_j} \right) \rho^T_{x_{n-1},\nu_{n-1}} \; = \; \frac{W(\widetilde{C}_k)}{W(C_k)} \; ,
\label{ratioworm3_new}
\end{equation}
where the last equality is now obvious from the definitions (\ref{rhos}). This concludes the proof of detailed balance for the
even-odd worm algorithm. We are currently exploring the performance of the even-odd worms in numerical tests in the $\phi^4$ 
theory.

\section{Canonical worldline simulations}

Let us now come to our second topic, the use of the worldline representation for canonical simulations, i.e., simulations at a fixed
net particle number. As we have discussed in Section 2, the net particle number $N$ corresponds to the temporal winding number 
$\omega[k]$ of the worldlines of conserved $k$-flux. Thus for every dual configuration we can identify the net particle number 
in a unique way. For simulating in a sector with a fixed net-particle number $N$ we can start with an initial configuration with a fixed net 
winding number $\omega[k] = N$, e.g., by placing $N$ simply winding temporal loops of $k$-flux, and then use a Monte Carlo update 
that does not change the winding number. This can either be implemented by reflecting the worms described in the previous section at the 
last timeslice or by offering local updates that change the flux around single plaquettes combined with offering flux that winds only 
around the spatial boundaries. This simulation at fixed $N$ corresponds to a simulation of the ensemble with a canonical 
partition sum $Z_N$.  An implementation of canonical worldline simulations based on these ideas was presented in 
\cite{canonical_phi4}. The numerical simulations were done for the 2-dimensional charged $\phi^4$ field and we here 
briefly summarize the key steps and some results.

The approach described in the previous paragraph allows one to calculate observables at fixed net particle numbers $N$ at a given
spatial extent $N_s$ and thus at a net particle density $n = N/N_s^{\, d-1}$ for the $d$-dimensional system. In principle 
one then can study these observables as a function of $n$ and in this way describe all finite density physics. 

Among the observables
one can define in the canonical ensemble is also the chemical potential which is defined as the derivative of the free energy with 
respect to the particle number density $n$,
\begin{equation}
\mu (n) \; = \; \frac{\partial f(n)}{\partial n} \; = \;  \frac{N_s}{2} 
\Bigg[ f \left(\frac{N+1}{N_s}\right) - f \left(\frac{N-1}{N_s}\right)\Bigg] + \mathcal{O}\left(\frac{1}{N^2_s}\right) \; ,
\label{mudef}
\end{equation} 
where $f(n) =  -  \ln Z_N / V$ at spatial extent $N_s$, such that $n = N/N_s^{\, d-1}$. In the last step of (\ref{mudef}) 
we have already discretized the derivative with respect to $n$ (this form is specific for $d = 2$  where 
$n = N/N_s$). 

The free energy $f(n) = f(N/N_s^{d-1})$ cannot be computed directly in a Monte Carlo simulation, but may be determined by integrating 
suitable observables. Here we use as observable the expectation value $\langle |\phi|^4\rangle$, i.e., 
the derivative of $f(n)$ with respect to the coupling $\lambda$. Integrating this differential equation we find,
\begin{equation}
f(n) \;\;  = \;\; f(n)\Big|_{\lambda=0} \; + \; \int_{0}^{\lambda} \! \!\! d \lambda^{\prime}  \; \Big\langle |\phi|^4\Big\rangle 
\Big|_{\lambda^{\prime}, N, N_s} \; .
\label{fintegral}
\end{equation}
As an integration constant appears $f(n) |_{\lambda=0}$, i.e., 
the free energy at vanishing coupling which can be evaluated with Fourier transformation.  The integral in (\ref{fintegral}) 
has to be done numerically. For that the expectation value $\langle |\phi|^4 \rangle |_{\lambda^{\prime}, N, N_s} $ 
has to be determined in canonical simulations at fixed $N$ and $N_s$ at several values $\lambda^\prime$ of the quartic coupling. 
We found \cite{canonical_phi4} that the behavior of $\langle |\phi|^4\rangle|_{\lambda^{\prime}, N, N_s}$ is rather smooth as a 
function of $\lambda^\prime$ and that the numerical integral in (\ref{fintegral}) can be determined accurately.

\begin{figure}
\begin{center}
\includegraphics[height=55mm,clip]{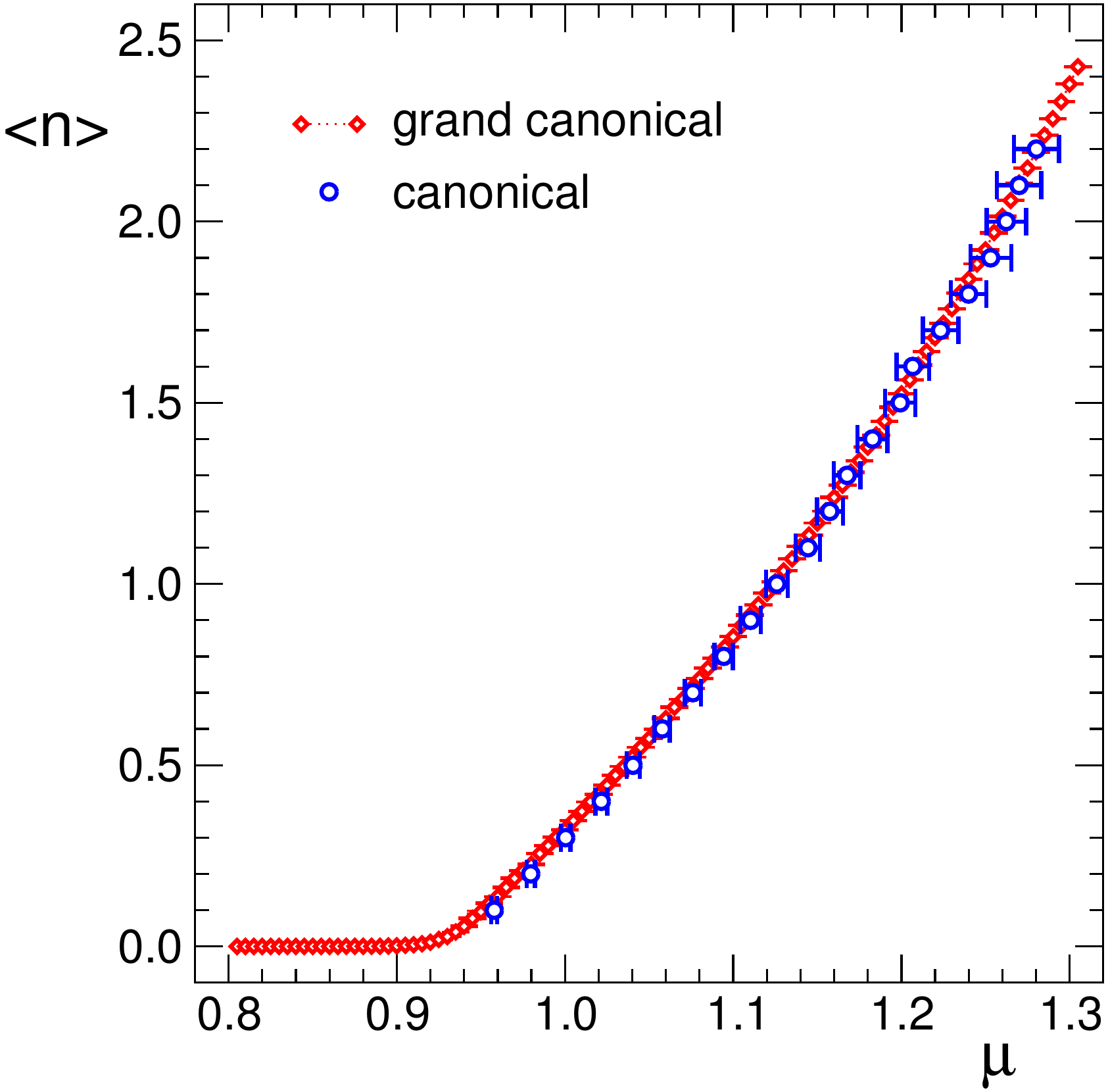}
\hspace{3mm}
\includegraphics[height=55mm,clip]{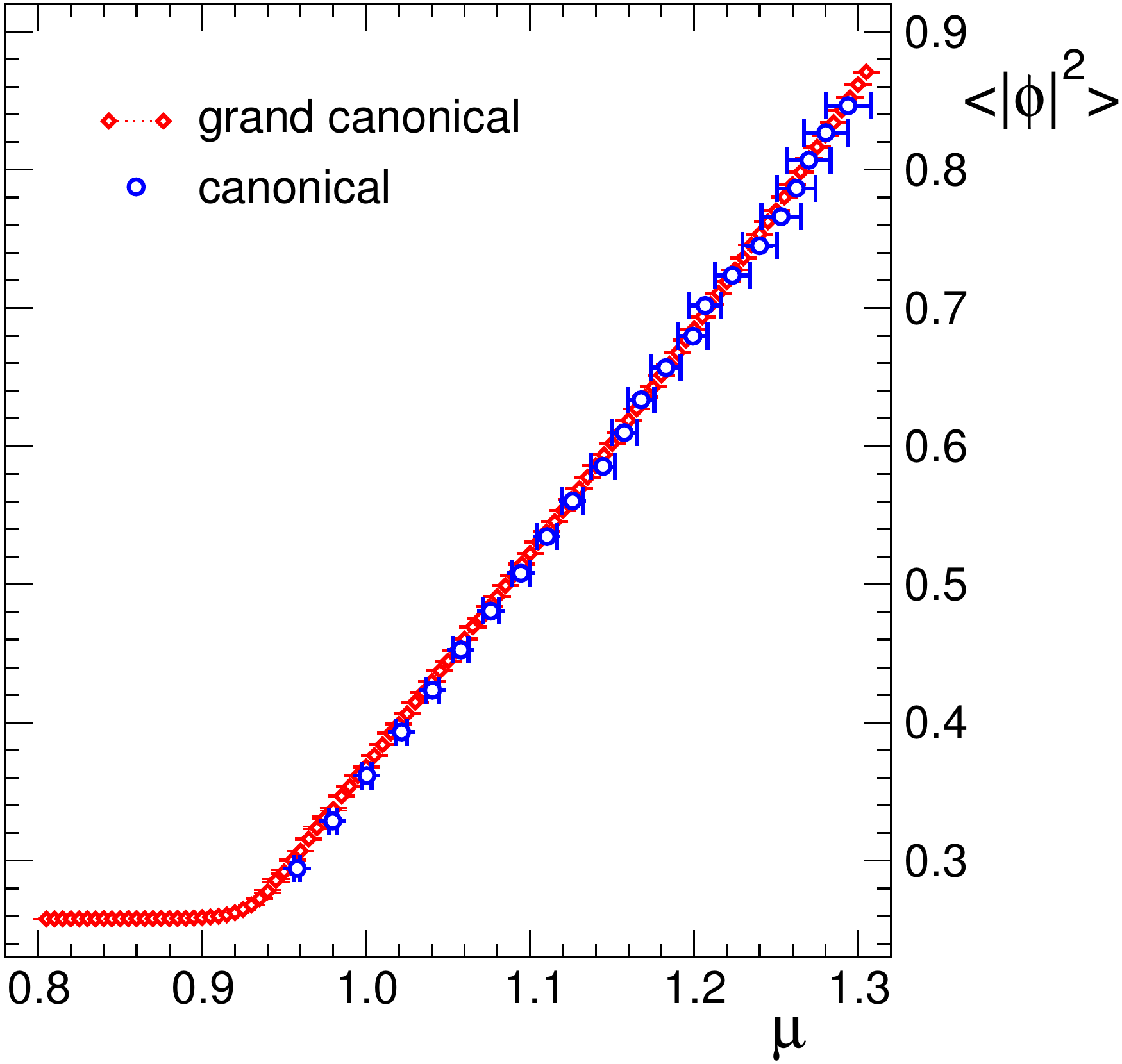}
\end{center}
\caption{The relation between the particle number density $\langle n \rangle$ and the chemical potential $\mu$ (lhs. plot) and the 
field expectation value $\langle |\phi|^2 \rangle$ versus $\mu$ (rhs.). We compare the grand canonical results to those determined from 
the canonical simulations. For the latter $\mu$ is an observable and thus we have horizontal error bars. The data are for
$d = 2$ dimensions with $\lambda = 1.0$,
$\eta = 4.01$, $N_s = 10$ and $N_t = 100$.}
\label{fig:compare}	
\end{figure}

In the lhs.~plot of Fig.~\ref{fig:compare} we show with blue circles our results for the relation between $n$ and $\mu$ from
the canonical determination based on (\ref{mudef}) and (\ref{fintegral}). Note that we plot $\mu$, which in the canonical
ensemble is an observable, on the horizontal axis, such that the canonical data points have horizontal error bars. 
The canonical data in both plots of Fig.~\ref{fig:compare} are from a simulation of the 2-dimensional 
canonical ensemble in the worldline form. The update was 
done with the local strategy of updating flux around plaquettes and spatially winding loops, using statistics of 
$10^5$ configurations separated by 10 combined sweeps for decorrelation and $5 \times 10^4$ sweeps for equilibration.
The couplings and lattice size are $\lambda = 1.0$, $\eta = 4.01$, with $N_s = 10$ and $N_t = 100$.

The canonical relation between $n$ and $\mu$ can now be compared to the corresponding grand canonical determination, where one
evaluates $\langle n \rangle$ as a function of $\mu$ using the dual representation (\ref{observables}). These are the data represented by
red diamonds in Fig.~\ref{fig:compare} which  we generated with the worm algorithm using a statistics of 
$4 \times 10^5$ configurations separated by 10 worms for decorrelation after  $2 \times 10^5$ worms for equilibration.
It is obvious that the two data sets nicely fall on top of each other and we conclude that with the canonical worldline approach we 
can reliably determine the relation between $n$ and $\mu$.

Once the relation between $n$ and $\mu$ is known we can also convert other observables from the canonical determination as 
a function of $n$ into their grand canonical form, i.e., express them as a function of the chemical potential. In the rhs.~plot of 
Fig.~\ref{fig:compare} we illustrate this for the field expectation value $\langle | \phi |^2 \rangle$.  Again we compare the canonical 
results with a direct determination in the grand canonical ensemble and also here find excellent agreement, such that we 
conclude that the canonical worldline approach provides an interesting alternative to grand canonical simulations. 

We have shown for the example of the 2-dimensional scalar $\phi^4$ field that the canonical worldline approach is a 
possible alternative for Monte Carlo simulations at finite density. However, for the bosonic example used here for
developing and testing the method the possible advantages of the canonical approach cannot be fully appreciated. We expect that 
for fermionic systems the canonical approach is considerably more interesting because there the Grassmann nature of the
matter fields leads to worldlines with more rigid constraints: Each site of the lattice has to be run through by a loop, be the endpoint
of a dimer or is occupied by a monomer. This additional constraint makes the system much more stiff in a Monte Carlo simulation 
and the insertion of additional winding loops, i.e., additional charges is rare. This has, e.g., been observed in the simulation 
\cite{schwinger_theory2,schwinger_simulation,daniel} of the worldline form \cite{schwinger_theory} of the massless Schwinger model.
It was found that in the grand canonical worldline simulation in particular the particle number suffers from very long autocorrelation 
times. A canonical worldline simulation as we have described here overcomes this problem, since the net-winding number of
the fermion loops and thus the particle number is kept fixed. First tests with simulations of the worldline form 
\cite{schwinger_theory} have shown that with the canonical worldline approach autocorrelation problems are much milder.

\section{Particle condensation and scattering data}

Let us now come to the third topic which we here study using the worldline representation of the charged 
$\phi^4$ field: The connection between scattering data of a theory and the condensation threshold for the two-particle sector 
in the grand canonical ensemble at low temperature.

\begin{figure}
\begin{center}
\includegraphics[height=55mm,clip]{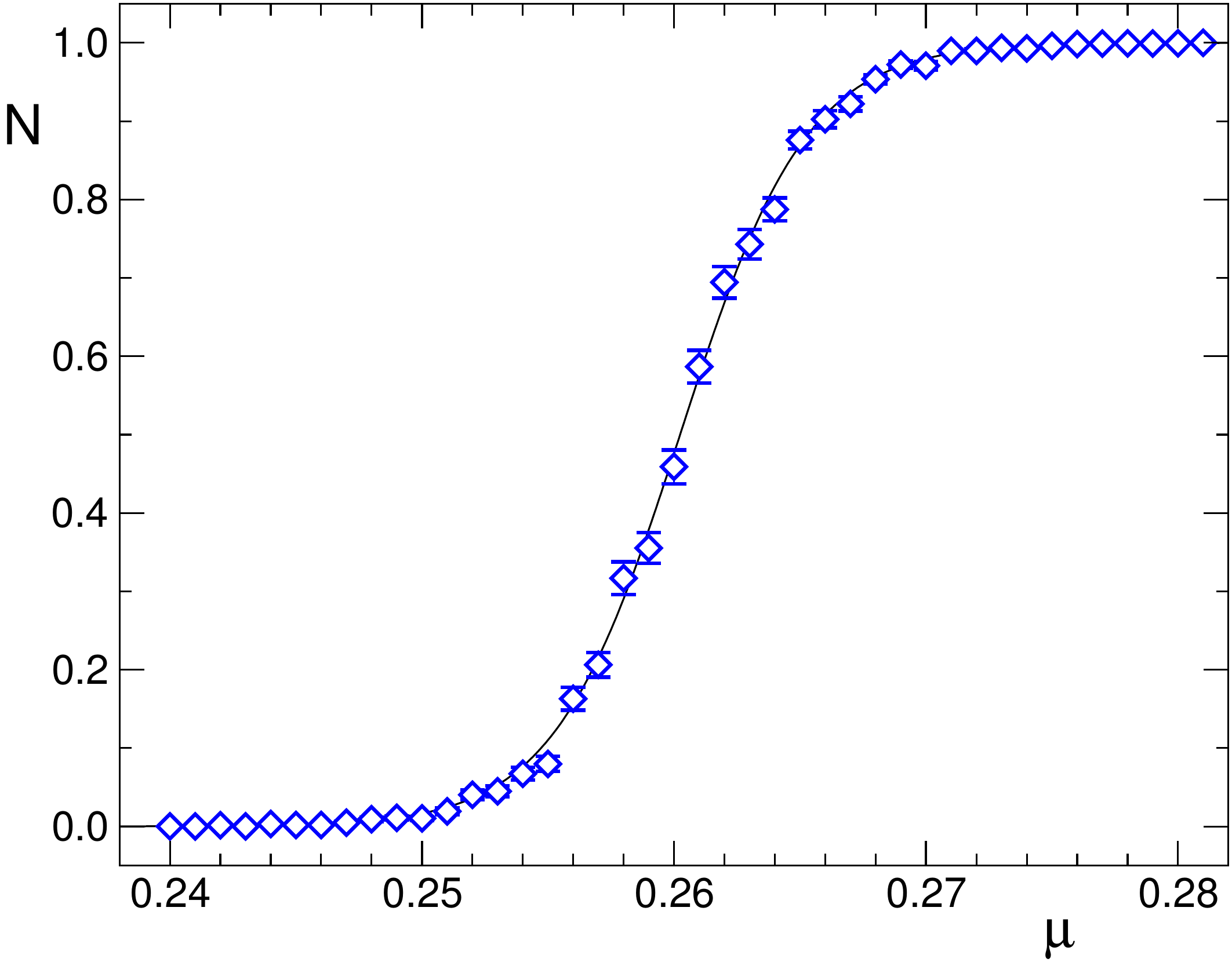}
\includegraphics[height=55mm,clip]{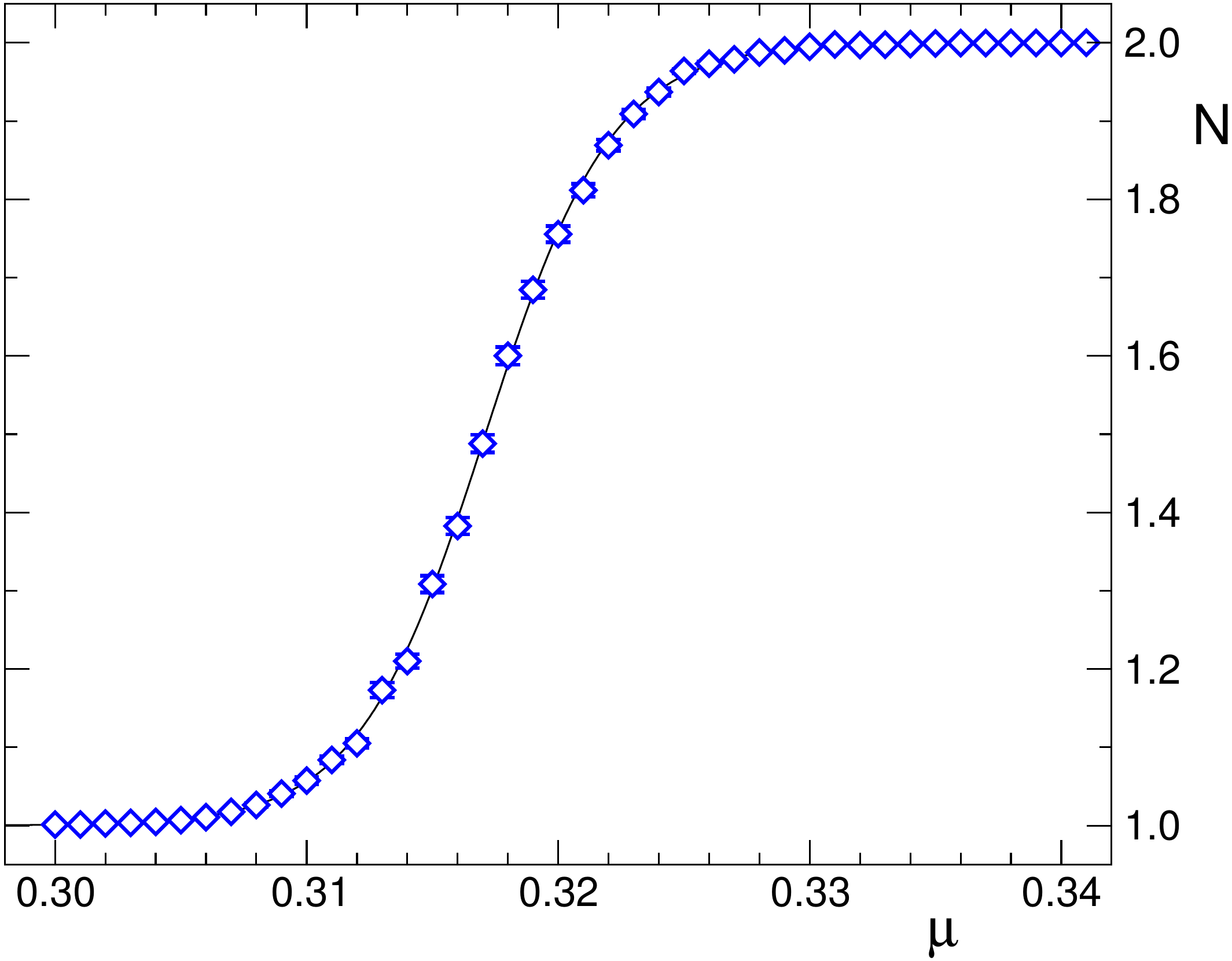}
\end{center}
\caption{Condensation for the 1-particle (lhs.~plot) and the 2-particle sectors (rhs.) in the 2-$d$ case. 
We show the particle number $N$ as a function of
the chemical potential $\mu$ for $\eta = 2.6, \lambda = 1.0$, $N_s = 16$ and a temporal extent of $N_t = 400$. The symbols
represent the numerical data from the worldline simulation and the curves the fit with the logistic function as discussed in the text.}
\label{fig:condensation}	
\end{figure}

The existence of a real and positive worldline representation allows one to study condensation of particles at very low temperature. 
To illustrate the condensation phenomena one can observe in such a study, in Fig.~\ref{fig:condensation} we show   
the expectation value of the net particle number $N$ as a function of the chemical potential $\mu$ for the 2-$d$ case. 
The grand canonical worldline data for 2-$d$ used in this section  
were generated with the worm algorithm with $4 \times 10^5$ configurations 
separated by 10 worms for decorrelation after $2 \times 10^5$ worms for equilibration. The chosen coupling parameters are 
$\eta = 2.6, \lambda = 1.0$ and a temporal extent of $N_t = 400$. The worldline data we use in this section for the 
4-dimensional case are based on $10^5$ to $2\times 10^5$ configurations 
generated at $\eta = 7.44, \lambda = 1.0$ using spatial extents of $N_t = 320$ and $N_t = 640$. 

In the lhs.\ plot of Fig.~\ref{fig:condensation} we show the results for $N$ in a range of $\mu$ 
where the net particle number rapidly increases from 
$N = 0$ to $N = 1$ at a critical chemical potential $\mu_c^{(1)} \sim 0.26$. In the rhs.\ plot we show the second step from 
$N = 1$ to $N = 2$ at a critical chemical potential $\mu_c^{(2)} \sim 0.32$. 
Although with $N_t = 400$ we work already at a relatively low temperature of $T = 1/400$ in lattice units we still observe rounding 
from temperature effects near the thresholds $\mu_c^{(1)}$ and $\mu_c^{(2)}$. Of course one can reduce the rounding by increasing 
$N_t$ further, which on the other hand drives up the numerical cost. In order to determine the critical values $\mu_c^{(i)}$ also at 
non-zero $T$ we fit the function $N(\mu)$ in the vicinity of the condensation steps with the logistic function
$N(\mu) = [1 +  \exp(-k[\mu-\mu_c^{(i)}])]^{-1} + i - 1$. As the plots in Fig.~\ref{fig:condensation} show, these fits describe 
the data near the two condensation thresholds $i = 1,2$ very 
well and the corresponding critical values $\mu_c^{(i)}$ are obtained as 
one of the two fit parameters.

\begin{figure}[p]
\begin{center}
\includegraphics[height=55.5mm,clip]{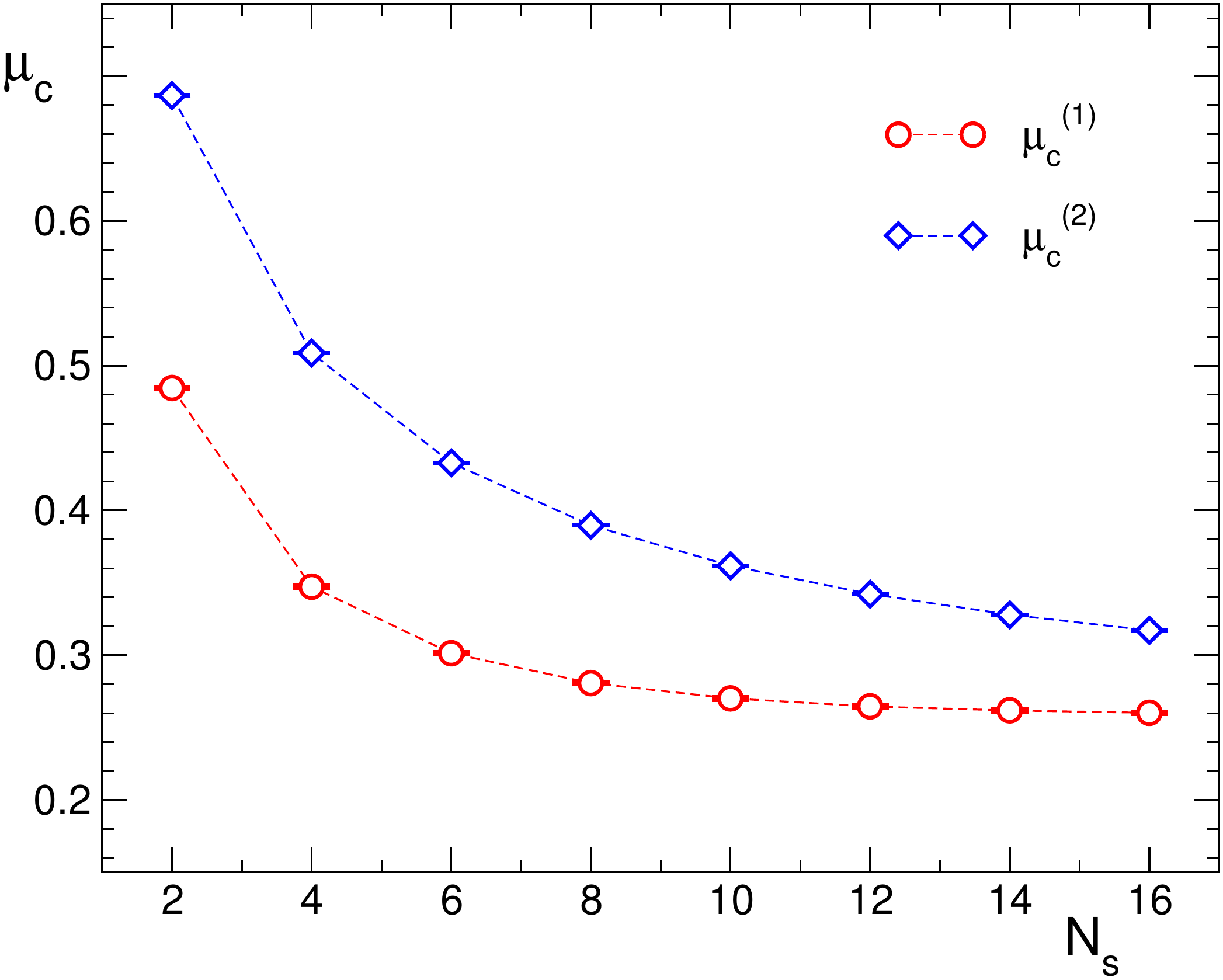}
\includegraphics[height=55.5mm,clip]{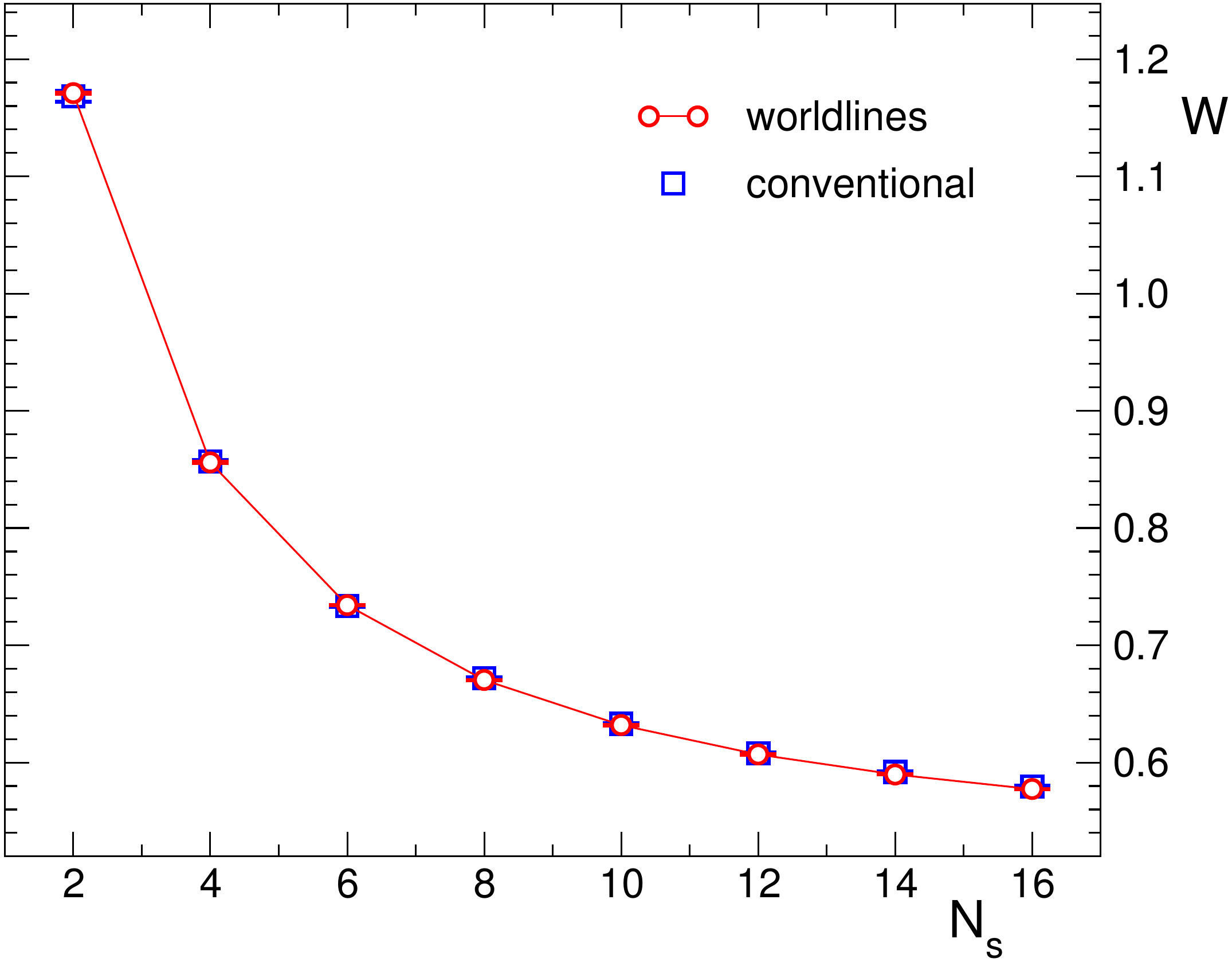}
\end{center}
\caption{Lhs.: Results for the critical chemical potential values $\mu^{(1)}_c$ and $\mu^{(2)}_c$ from the 2-$d$ worldline simulation
as a function of the spatial lattice extent $N_s$. 
Rhs.: Comparison of the results for 2-particle energy from the worldline simulation (circles) and the conventional 
determination with 4-point functions (squares). For both plots the parameters are $\eta = 2.6$ and $\lambda = 1.0$. For the 
worldline data $N_t = 400$ was used and for the conventional 4-point functions $N_t = 32$ and $N_t = 64$.}
\label{fig:W2results2d}	
%\end{figure}
\vskip10mm
%\begin{figure}
\begin{center}
\includegraphics[height=56mm,clip]{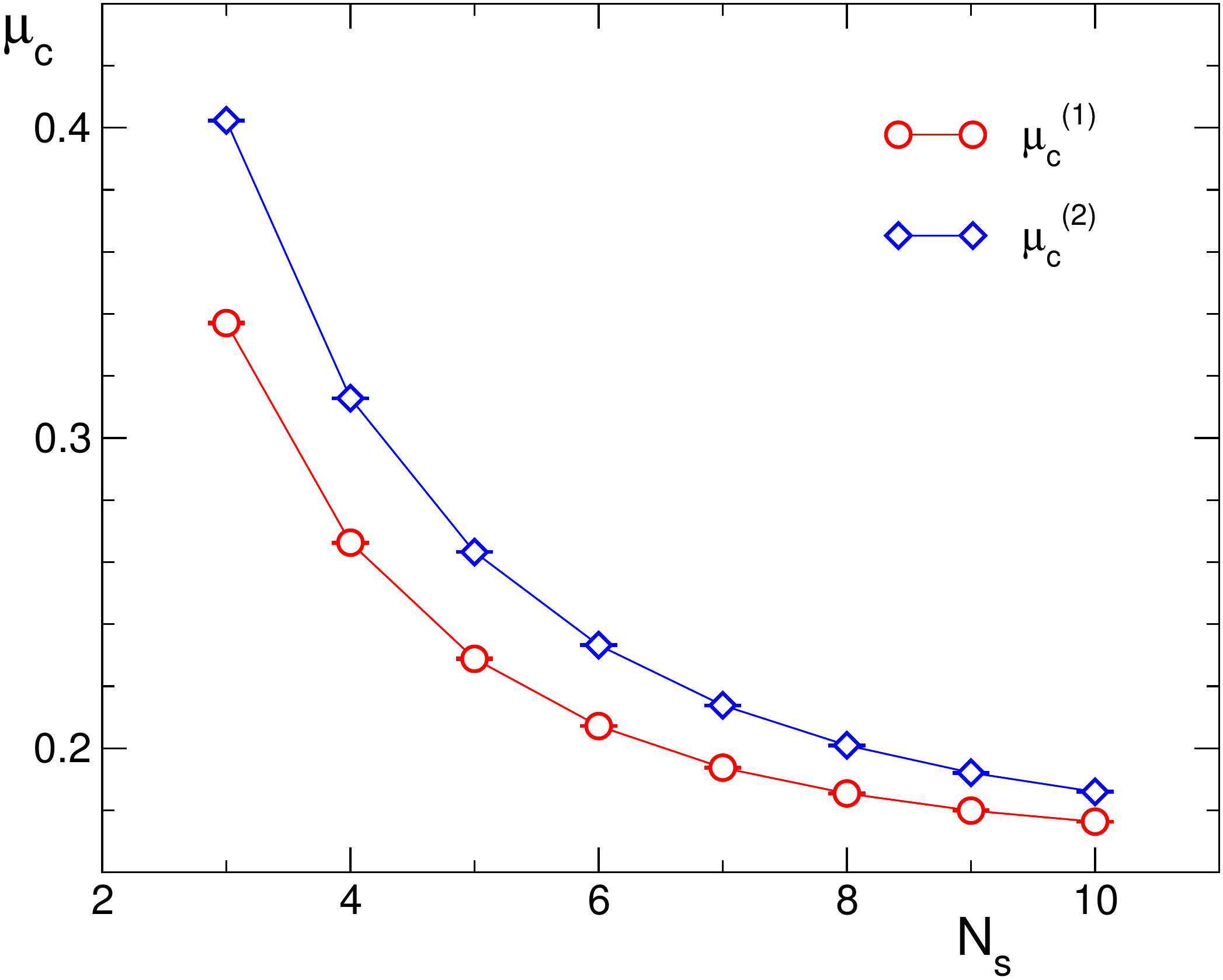}
\includegraphics[height=56mm,clip]{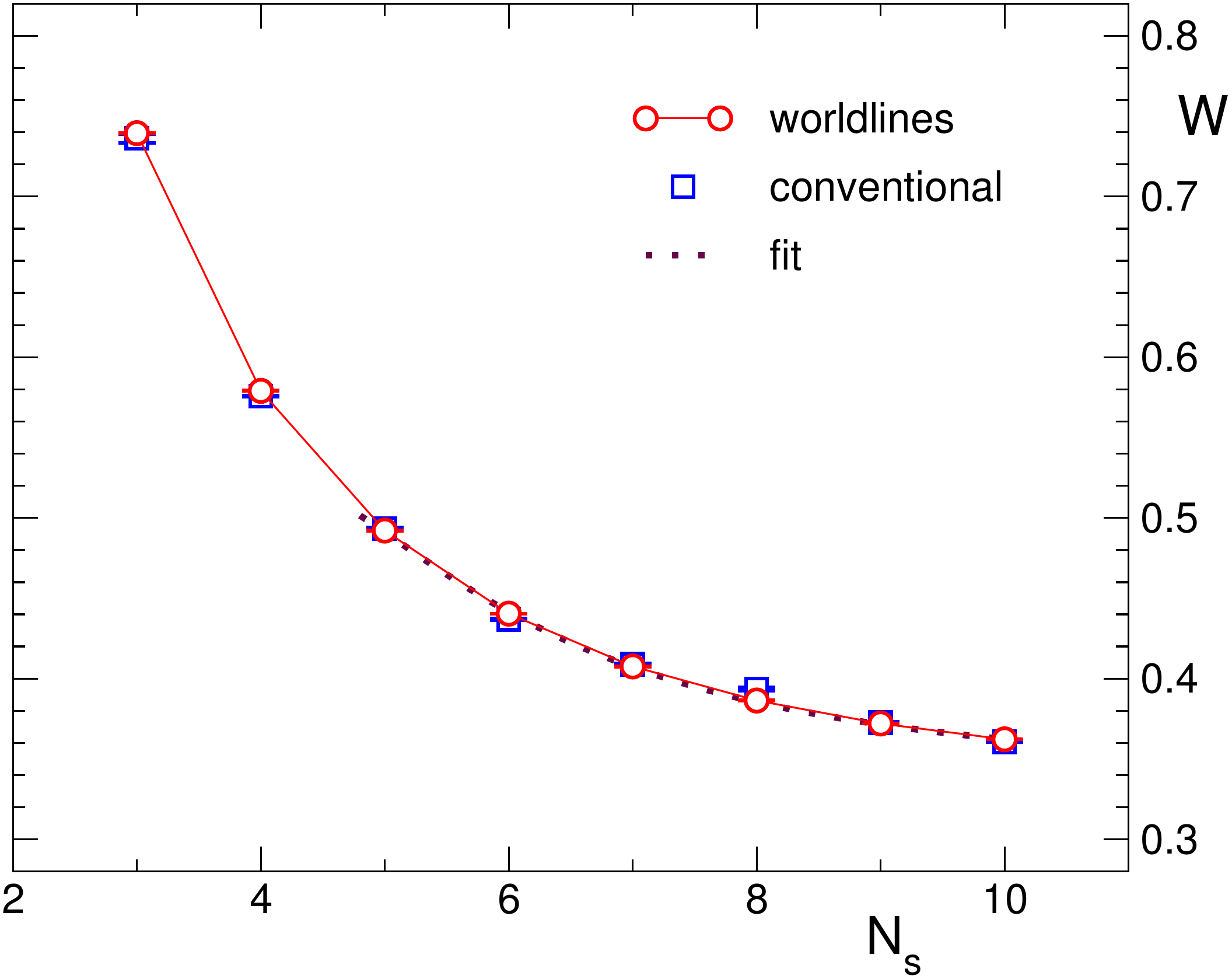}
\end{center}
\caption{Lhs.: Results for the critical values of the chemical potentials as a function of the spatial lattice extent 
$N_s$ determined from the worldline simulation of the 4-dimensional case. 
Rhs.: Comparison of the different determinations of the 2-particle energy. The results from the condensation thresholds in the 
worldline approach are shown as circles, the results from the analysis of 4-point functions in a conventional simulation are
represented by squares. For both plots the parameters are $\eta = 7.44$ and $\lambda = 1.0$, with $N_t$ between 
$N_t = 320$ and $N_t = 640$ for the worldline data and $N_t = 48$ and $N_t = 64$ for the conventional 4-point functions.
}
\label{fig:W2results4d}	
\end{figure}

The position $\mu_c^{(1)}$ of the first condensation threshold coincides with the renormalized physical mass
of the system, i.e., $\mu_c^{(1)} = m_{phys}$. However, also the second threshold can be related to a specific energy value:
In \cite{scattering_O3} it was noted that in the limit of vanishing temperature the critical value $\mu_c^{(2)}$ coincides with 
$W - m_{phys}$, where $W$ is the 2-particle energy. The relations that connect the condensation thresholds to physical 
quantities thus read \cite{scattering_O3}  
\begin{equation}
m_{phys} \; = \; \mu_c^{(1)}  \qquad \mbox{and} \qquad
W \; = \; \mu_c^{(1)} \, + \, \mu_c^{(2)} \; .
\end{equation}
The results for $\mu_c^{(1)}$ and $\mu_c^{(2)}$ and thus $m_{phys}$ and $W$ depend on the 
spatial extent $N_s$, as can be seen from the lhs.\ plot of 
Fig.~\ref{fig:W2results2d} for the 2-$d$ case and the lhs.\ plot of Fig.~\ref{fig:W2results4d} for 4-$d$. In both plots we show the results for 
$\mu_c^{(1)}(N_s)$ and $\mu_c^{(2)}(N_s)$  as determined from the corresponding condensation thresholds.

The physical mechanisms for the finite volume effects of $m_{phys}$ and $W$ were discussed in two papers by L\"uscher 
\cite{Luscher:1985dn,Luscher:1986pf} where also the finite volume scaling formulas for $m_{phys}(N_s)$ and $W(N_s)$ are given. 
In particular the second paper \cite{Luscher:1986pf} analyzes the 2-particle energy $W$ and relates it to scattering data:
From the finite volume scaling of $W(N_s)$
one can extract the scattering length $a_0$ and for the 2-dimensional case even the full scattering 
phase shift $\delta(k)$ (see also \cite{Luscher:1990ck}). 

Before we start with the determination of scattering data from the 2-particle energies, we discuss an important cross-check of the 
results for $W$ obtained from the condensation thresholds. The 2-particle energies may also be computed from 4-point
functions in the conventional formulation: Denoting by $\widetilde{\phi}(t)$ the sum of the $\phi_{x}$ over the time slice
at $t$, i.e., the zero-momentum projected field, we can obtain $W$ from the exponential decay of the 4-point function
\begin{equation}
\left\langle \, \widetilde{\phi}(t) \, \widetilde{\phi}(t) \; \widetilde{\phi}(0)^\star \widetilde{\phi}(0)^\star  \right \rangle \; \propto  \; e^{- W t} \; .
\label{4-point}
\end{equation}
Certainly one may improve the determination of $W$ by considering a full correlation matrix of 4-point correlators, but we found that the
simple correlator (\ref{4-point}) entirely dominates the signal. We implemented the determination of $W$ based on (\ref{4-point}) for
both the 2-$d$ and the 4-$d$ cases for the same values of $N_s$ as studied for the condensation thresholds. 
The corresponding data points are shown as squares in the rhs.\ plots of Fig.~\ref{fig:W2results2d} and Fig.~\ref{fig:W2results4d}.
We find excellent agreement of the results from the two determinations and conclude that our interpretation of the 2-particle 
threshold as  $\mu_c^{(2)} \; = \; W - \mu_c^{(1)}$ applies and that our determination of the critical values $\mu_c^{(i)}$ at 
non-zero temperature from the fit with the logistic function  
is reliable. The 2-dimensional conventional data for this comparison were computed from 
$10^6$ configurations separated by 10 sweeps of local Monte Carlo updates for decorrelation after $10^4$ sweeps for equilibration using
temporal lattice extents of $N_t = 32$ and $N_t = 64$.
In 4 dimensions we used $5\times 10^5$ configurations and $N_t = 48$ and $N_t =  64$.

Having confirmed the relation of the second condensation threshold to the 2-particle energy with the comparison to the results 
from 4-point functions in the conventional approach let us now come to the determination of the scattering data.
We begin the analysis with the 2-dimensional case (compare \cite{Luscher:1990ck}). 
There the 2-particle energy $W$ is related to the relative momentum 
$k$ of the two particles via
\begin{equation}
W \; = \;  2\sqrt{ \,  \left(\mu^{(1)}_{c}\right)^2 + k^2 \,} \; .
\label{scatter2da}
\end{equation}
This equation may be inverted to determine the momentum $k$. Due to the finite volume the momentum $k$ is subject to the 
quantization condition, 
\begin{equation}
e^{2i\delta(k)} \; = \; e^{-i k N_s} \; ,
\label{scatter2db}
\end{equation}
which also contains the phase shift $\delta(k)$ for that momentum. Thus by combining (\ref{scatter2da}) and (\ref{scatter2db})
one may extract the scattering phase shift $\delta(k)$ from the 2-particle energies $W(N_s)$. Varying $N_s$ allows one to access
different values of the relative momentum $k$, such that $\delta(k)$ can be determined for a whole range of momenta. 

\begin{figure}
\begin{center}
\includegraphics[height=60mm,clip]{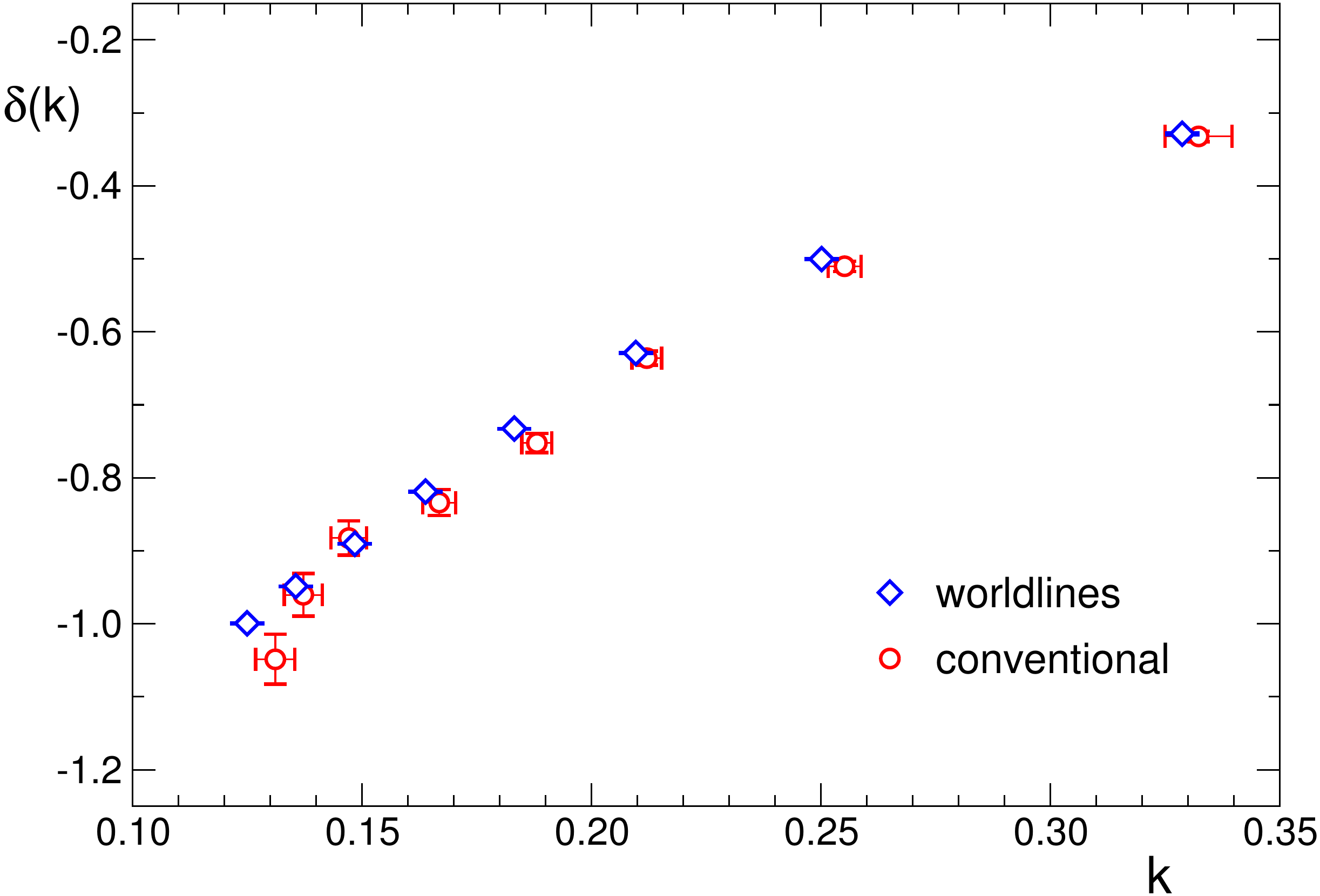}
\end{center}
\caption{Scattering phase shift $\delta(k)$ for 2-$d$ 
as a function of the momentum $k$. We compare the results from the worldline calculation to the data
from the conventional approach. The parameters of the simulations are the same as in Fig.~4.
}
\label{phaseshift2d}	
\end{figure}

In Fig.~\ref{phaseshift2d} we show the results for this analysis of the 2-dimensional case (compare also the results
for the 2-$d$ Ising model  \cite{ising_scattering1,ising_scattering2}
which corresponds to a particular limit of the $\phi^4$ system studied here). We plot the results for $\delta(k)$ 
as a function of $k$ and again compare the data based on the condensation thresholds in the worldline simulation to the results
from analyzing 4-point functions in the conventional approach. We find very good agreement of the two data sets, and point out that 
the results from the worldline simulation scatter less and have smaller error bars.

We conclude the discussion of the relation between the 2-particle condensation threshold and scattering data with the analysis of the
4-dimensional case. There the 2-particle energy can be expanded in a power series in $a_0/N_s$ \cite{Luscher:1986pf},
\begin{equation}
W(N_s) \; = \; 2 m_0   \; + \; \frac{4 \pi \, a_0}{m N_s^3} \left[ 1 + c_1 \frac{a_0}{N_s}  + c_2 \left( \frac{a_0}{N_s} \right)^2
+ c_3 \left( \frac{a_0}{N_s} \right)^3 \; ...  \right] \; .
\label{W2luscher}
\end{equation}
Here $m_0$ is the physical mass in lattice units in the infinite volume limit, i.e., $m_0 = \lim_{N_s \rightarrow \infty} \, \mu_c^{(1)}(N_s)$, 
and $a_0$ the scattering length in lattice units. The coefficients $c_1$ and $c_2$ were determined in \cite{Luscher:1986pf} with values  
$c_1 =  - 2.837297$ and  $c_2 = 6.375183$, and $c_3$ we use as an additional fit parameter 
that allows us to use our data for $W(N_s)$ also for
relatively small $N_s$. The physical mass in the infinite volume limit $m_0$ was determined from fitting the data for $\mu_c^{(1)}(N_s)$ 
with the 2-parameter form $\mu_c^{(1)}(N_s) = m_0 + c N_s^{\,-3/2} e^{-m_0 N_s}$ \cite{Rummukainen:1995vs}. For the parameters
we use ($\lambda = 1.0, \eta = 7.44$), this gives a value of $m_0 =  0.168(1)$. Using this as input in (\ref{W2luscher}) reduces the
fit of $W(N_s)$ to a 2-parameter fit with parameters $a_0$ and $c_3$, which we applied to the worldline data for $W(N_s)$
in the range from $N_s = 5$ to $N_s = 10$. This fit is shown as a dashed line in the rhs.\ plot of Fig.~\ref{fig:W2results4d}. 

Our final results for the physical mass $m_0$ in lattice units determined from fitting the data for $\mu_c^{(1)}(N_s)$ and 
the scattering length $a_0$ in lattice units determined from (\ref{W2luscher}) are:
\begin{equation}
m_0 \; = \; 0.168(1) \qquad  ,  \qquad  a_0 \; = \; - \, 0.32(2) \; .
\label{results}
\end{equation}
The error we quote for $m_0$ is the statistical error, while for $a_0$ we tried to estimate the systematic error by adding an additional term
for higher corrections in (\ref{W2luscher}) and also using $m_0$ as an additional independent fit parameter (this gives a value of
$m_0 = 0.169$ which is consistent with the independent determination from $\mu_c^{(1)}$ quoted in (\ref{results}). 
We stress that these results are preliminary and we are currently improving the analysis and generate data for a second lattice spacing.

To summarize: Following the analysis in  \cite{scattering_O3} we have shown that the second condensation threshold $\mu_c^{(2)}$ 
at low temperatures is related to the 2-particle energy $W$. We cross-checked this relation for both, the 2- and the 
4-dimensional cases with a determination of $W$ from 4-point functions computed in a simulation in the conventional 
representation of the $\phi^4$ lattice field theory. In a subsequent analysis of the finite volume behavior using L\"uscher's formulas
we determined the scattering phase shift $\delta(k)$ for the 2-dimensional case and the scattering length $a_0$ in 4 dimensions.

Also here the findings discussed for the $\phi^4$ lattice field theory clearly go beyond that simple model. A particularly 
simple and interesting test case would be QCD with two degenerate quark flavors. In this case the introduction of an isospin chemical 
potential $\mu_I$ does not lead to a complex action problem and $\mu_I$ could be used to condense pions. The second
condensation threshold could then be analyzed along the lines discussed here to extract parameters for pion scattering.

%----------------------------------------------------------------------------

\section{Summary}

In this contribution we have presented three developments for Monte Carlo simulations with worldlines, using the charged $\phi^4$ field
as the lattice field theory to illustrate these ideas. The three developments are: 1) New algorithmic ideas for worm updates. 
2) The possibility and perspectives for canonical worldline simulations. 3)  The relation between condensation at low temperature 
and scattering data.

Concerning the new strategies for worm algorithms the key issue is to find efficient algorithms for worldline models which in addition to the 
usual weights on the links also have weights on the sites of the lattice. For the $\phi^4$ case these weights come from the radial 
degrees of freedom, but in other models may also 
origin from, e.g., Haar measure contributions. The site weights couple the flux variables on all links
attached to a site and imply that the Metropolis ratios for the starting- or the closing step of the worms can be very small. We here
discuss two possibilities for eliminating or mildening this problem: The introduction of an amplitude factor that controls the starting- 
and closing probabilities to optimize the length of the worms. A second technique we discuss is an 
even-odd staggered worm, where one chooses different acceptance probabilities for even to odd and odd to even steps.  

The second issue we address is the use of the worldline formulation for canonical simulations. Since in the worldline representation the 
net-particle number is the temporal winding number $\omega$ of the worldlines, we can use an update that does not change $\omega$
and in this way simulate the canonical ensemble. We explore the feasibility of such a simulation by cross-checking also 
with grand canonical results and establish that canonical worldline simulations are an interesting alternative for finite density simulations.

Finally we explore the relation of the critical chemical potential values $\mu_c^{(1)}$ and $\mu_c^{(2)}$ for the onset of
1- and 2-particle condensation at low temperatures to scattering data. 
The 2-particle energy $W$ is given by the sum  $W = \mu_c^{(1)} + \mu_c^{(2)}$ 
and using grand canonical worldline simulations we determine the critical chemical potentials as a function of the spatial lattice extent  
$N_s$ and thus obtain $W(N_s)$. Fitting $W(N_s)$ with L\"uscher's formula we obtain the scattering phase shift for the 
2-$d$ case and the scattering length in 4 dimensions. Our results are cross-checked with a determination of $W(N_s)$ from 4-point
functions in the conventional formulation, and we thus establish that the low-temperature condensation thresholds are related to 
scattering data.

We conclude with stressing that all three topics we have discussed here go beyond the simple $\phi^4$ theory we have used as the 
physical system for our presentation. The further development of the methods and concepts discussed in this contribution is currently
pursued also in the context of other lattice field theories.

\vskip5mm
\noindent
\textbf{Acknowledgments:}  We thank Daniel G\"oschl and Thomas Kloiber for interesting discussions.
This work is supported by the FWF DK W 1203, "Hadrons in Vacuum, Nuclei and Stars", 
and by the FWF project I 2886-N27 in cooperation with DFG TR55, ''Hadron Properties from Lattice QCD''. 
%\clearpage
\vskip5mm

\bibliography{lattice2017}

%%%%%%%%%%%%%%%%%%%%%%%%%%%%%%%%%%%%%%%%%%%%%%%%%%%%%%%%%%%%%%%%%%%%%%%%%%%%%
\end{document}